\documentclass{emulateapj}

\slugcomment{Accepted for publication in \apj}

\shorttitle{Parsec-scale jets in radio-loud NLS1s}
\shortauthors{Doi, Asada, \& Nagai}

\begin{document}

\title{Very Long Baseline Array Imaging of Parsec-scale Jet Structures\\ in Radio-loud Narrow-line Seyfert 1 Galaxies\\}

\author{Akihiro~Doi\altaffilmark{1,2}, Keiichi~Asada\altaffilmark{3}, and Hiroshi~Nagai\altaffilmark{4}}

\altaffiltext{1}{The Institute of Space and Astronautical Science, Japan Aerospace Exploration Agency, 3-1-1 Yoshinodai, Chuou-ku, Sagamihara, Kanagawa 252-5210, Japan}\email{akihiro.doi@vsop.isas.jaxa.jp}
\altaffiltext{2}{Department of Space and Astronautical Science, The Graduate University for Advanced Studies,\\ 3-1-1 Yoshinodai, Chuou-ku, Sagamihara, Kanagawa 252-5210, Japan}
\altaffiltext{3}{Academia Sinica Institute of Astronomy and Astrophysics, P.O. Box 23-141, Taipei 10617, Taiwan}
\altaffiltext{4}{National Astronomical Observatory of Japan, 2-21-1 Osawa, Mitaka, Tokyo 181-8588, Japan}

\begin{abstract}
We conducted very long baseline interferometry~(VLBI) observations of five radio-loud narrow-line Seyfert~1~(NLS1) galaxies in milliarcsecond resolutions at 1.7~GHz~($\lambda$18~cm) using the Very Long Baseline Array~(VLBA).  Significant parsec-scale structures were revealed for three out of the five sources with high brightness temperature by direct imaging; this is convincing evidence for nonthermal jets.  FBQS~J1644+2619 with an inverted spectrum showed a prominent one-sided linear structure, indicating Doppler beaming with an intrinsic jet speed of $>0.74c$.  FBQS~J1629+4007, also with an inverted spectrum, showed rapid flux variability, indicating Doppler beaming with an intrinsic jet speed of $>0.88c$.  Thus, we found convincing evidence that these two NLS1s can generate at least mildly or highly relativistic jets, which may make them apparently radio loud even if they are intrinsically radio quiet.  On the other hand, the other three NLS1s had steep spectra and two of them showed significantly diffuse pc-scale structures, which were unlikely to be strongly beamed.  Thus, some NLS1s have the ability to generate jets strong enough to make them intrinsically radio loud without Doppler beaming.  NLS1s as a class show a number of extreme properties and radio-loud ones are very rare.  We build on these radio results to understand that the central engines of radio-loud NLS1s are essentially the same as that of other radio-loud AGNs in terms of the formation of nonthermal jets.  
\end{abstract}

\keywords{galaxies: active --- galaxies: jets --- galaxies: Seyfert --- radio continuum: galaxies --- techniques: interferometric}

\defcitealias{Doi_etal.2006a}{Paper~I}
\defcitealias{Doi_etal.2007}{Paper~II}
@
\section{INTRODUCTION}\label{section:introduction}
Active galactic nuclei~(AGNs) are believed to be powered by gravitational energy released from accreting matter onto supermassive black holes~(SMBHs) at the center of galaxies.  Narrow-line Seyfert~1 galaxies~(NLS1s) are a subclass of AGNs and are defined by the following optical spectral properties: 
(1)~permitted lines are slightly broader than forbidden lines;  
(2)~the full width at half-maximum~(FWHM) of the H$\beta$ emission line is $<$2000~km~s$^{-1}$; and 
(3)~the flux ratio [\ion{O}{3}]/H$\beta<$3, but exceptions are allowed when strong high-ionization iron lines such as [\ion{Fe}{7}] $\lambda$6087 and [\ion{Fe}{10}] $\lambda$6375, which are seldom seen in type-2 AGNs, are present~\citep{Osterbrock&Pogge1985,Pogge2000}.  
Other extreme properties of NLS1s are strong permitted Fe $_\mathrm{II}$ emission lines in optical and ultraviolet spectra \citep{Boroson&Green1992}, rapid and large-amplitude X-ray variability \citep{Leighly1999a}, a steep soft X-ray spectrum \citep{Boller_etal.1996,Brandt_etal.1997,Grupe_etal.1998,Grupe_etal.2004}, and a frequently observed blueshifted line profile \citep{Zamanov_etal.2002,Boroson2005,Leighly&Moore2004}.   In terms of radio luminosity ($10^{20}$--$10^{23}$~W~Hz$^{-1}$) and the compact source size ($\lesssim300$~pc; \citealt{Ulvestad_etal.1995,Moran2000}), the radio continuum properties of NLS1s and normal Seyfert~1 galaxies are not remarkably different.  NLS1s are generally radio-quiet objects\footnote{Radio loudness, $R$, is conventionally defined as the ratio of the 5-GHz radio to {\it B}-band flux densities, with a threshold of $R=10$ separating the radio-loud and radio-quiet objects \citep{Visnovsky_etal.1992,Stocke_etal.1992,Kellermann_etal.1994}.}.  The fraction of radio-loud objects in NLS1s is low \citep{Stepanian_etal.2003,Greene_etal.2006,Zhou_etal.2006}, that is, $\sim7$\% ($R>10$) and $\sim2.5$\% ($R>100$) (\citealt{Komossa_etal.2006a}, see also \citealt{Zhou&Wang2002,Yuan_etal.2008}), which is significantly lower than normal Seyfert galaxies and quasars \citep[$\sim10$\%--15\%;][]{Ivezic_etal.2002}.

There is increasing evidence that the unusual optical/UV/X-ray properties of NLS1s are physically related to the high mass accretion rates close to the Eddington limit \citep{Boroson&Green1992,Pounds_etal.1995,Sulentic_etal.2000,Boroson2002} of the relatively low-mass black holes~($\sim 10^5$--10$^7\ M_\sun$), which are located at the lower-mass end of the known population of SMBHs \citep[][and references therein]{Peterson_etal.2000,Grupe&Mathur2004,Komossa&Xu2007}.  Hence, NLS1s are expected to be promising samples for understanding the nature of the extreme accretion onto rapidly growing SMBHs.  The radio emission of AGNs and X-ray binaries is suggested to be also related to the black hole mass and accretion rate \citep{Merloni_etal.2003,Heinz&Sunyaev2003,Falcke_etal.2004}.  Many studies using large samples show a correlation between radio loudness and the black hole mass \citep{Laor2000,Lacy_etal.2001,Dunlop_etal.2003,McLure&Jarvis2004,Metcalf&Magliocchetti2006,Woo&Urry2002} and an anti-correlation between radio loudness and the accretion rate \citep{Ho2002,Lacy_etal.2001,Maccarone_etal.2003,Greene_etal.2006}.  These correlations predict that AGNs with high accretion rates and low-mass black holes such as NLS1s should be radio quiet.  In fact, as mentioned above, most NLS1s have been detected as radio-quiet objects.  In the analogy of X-ray binaries, NLS1s are at the high/soft state or at the very high state \citep{Mineshige_etal.2000,Wang&Netzer2003}, where radio emission is quenched \citep{Maccarone_etal.2003}.  

Radio-loud NLS1s are rare, but they do exist \citep{Siebert_etal.1999,Grupe_etal.2000,Zhou&Wang2002,Zhou_etal.2003,Whalen_etal.2006,Komossa_etal.2006a,Komossa_etal.2006b,Yuan_etal.2008}.  The radio powers of the radio-loud NLS1s exceed those of the most luminous starburst galaxies, suggesting that strong AGN jet emissions are expected to be responsible for making these NLS1s radio loud.  Although the black hole masses of the radio-loud NLS1s ($\sim 10^7\ M_\sun$) may be slightly larger than those of the radio-quiet NLS1s, the radio-loud NLS1s are still at a lower mass regime in AGNs and at a very high accretion rate regime (\citealt{Komossa_etal.2006a}, see also \citealt{Whalen_etal.2006,Yuan_etal.2008}); although such properties tend toward radio-quietness.  One possible explanation of the existence of radio-loud NLS1s is that the Doppler beaming effect on nonthermal jets influences radio loudness as well as some other radio-loud AGNs.  However, most radio-loud NLS1s show steep soft-excess X-ray spectra, which are the same as those of the radio-quiet NLS1s \citep{Komossa_etal.2006a,Yuan_etal.2008} and in contrast to the flat X-ray spectra due to the inverse Compton process in relativistic jets in radio-loud quasars~\citep{Reeves_etal.1997}.  The steep X-ray spectra of radio-loud NLS1s may be synchrotron tails originating from the relativistic jets \citep{Zhou_etal.2007,Yuan_etal.2008}, as seen in high-energy-peaked BL~Lac objects~\citep{Padovani&Giommi1995}.  On the other hand, \citet{Komossa_etal.2006a} indicated that many radio-loud NLS1s show steep radio spectra where strong beaming is generally not expected, although {\it very} radio-loud NLS1s may systematically show flat radio spectra \citep{Yuan_etal.2008}.  Several radio-loud NLS1s show physically unrealistic, high brightness temperatures, derived from radio flux variability.  This indicates the Doppler beaming effect if not interstellar scattering \citep{Zhou_etal.2003,Yuan_etal.2008}.  However, the Doppler beaming effect on the jets of NLS1s is not yet well established.  

Very long baseline interferometry~(VLBI) is one of the most powerful tools for revealing the radio properties in parsec~(pc) scales by direct imaging at milliarcsecond~(mas) resolutions.  Arcsecond-resolution observations have resolved the structures of only a few NLS1s because they are generally quite compact radio sources \citep{Ulvestad_etal.1995,Moran2000}.  VLBI observations on a large number of radio-quiet and radio-loud NLS1s are crucial for understanding the nature of the jet phenomena in these central engines.  VLBI images of only several objects have been reported; that is, for radio-quiet NLS1s: NGC~4395 \citep{Wrobel_etal.2001}, MRK~766, AKN~564 \citep{Lal_etal.2004}, and NGC~5506 \citep{Middelberg_etal.2004} and for radio-loud NLS1s: \object{SDSS J094857.31+002225.4} (\citealt{Doi_etal.2006a}, hereafter \citetalias{Doi_etal.2006a}), J16290+4007, J16333+4718, and B3~1702+457 \citep{Gu&Chen2010}.  SDSS~J094857.31+002225.4, which is the most radio-loud narrow-line object with $R\approx2000$, is suggested to be Doppler beamed by radio flux variability \citep{Zhou_etal.2003}, and it has been revealed that Doppler-boosted jets are required to explain the directly measured high brightness temperatures of its very compact radio emissions seen in the VLBI images \citepalias{Doi_etal.2006a}.  The recent discovery of $\gamma$-ray emission from SDSS~J094857.31+002225.4 by the {\it Fermi Gamma-Ray Space Telescope} \citep{Abdo_etal.2009a} and the subsequent VLBI observations have established the presence of a relativistic jet in this extreme narrow-line quasar \citep{Abdo_etal.2009b,Giroletti_etal.2011}.

We conducted VLBI observations for five radio-loud NLS1s at 1.7 and 8.4~GHz using the Very Long Baseline Array~(VLBA) and the Japanese VLBI Network~(JVN), respectively.  The results of the JVN observations, which have been published (\citealt{Doi_etal.2007}, hereafter \citetalias{Doi_etal.2007}), showed no significant radio structure in these five objects.  In the present paper, we report the observations at 1.7~GHz, where synchrotron radio jet structures are generally expected to be more prominent than at higher frequencies.  Based on the VLBA observations at 5~GHz for three radio-loud NLS1s that are three of our five samples~(for one source also at 2.3 and 8.4 GHz), \citet{Gu&Chen2010} reported that compact structures predominate radio emissions.    

The present paper is structured as follows.  The sample selection is shown in Section~\ref{section:sample}.  Section~\ref{section:observation} describes our observations and the data reduction procedures.  In Section~\ref{section:result}, we present the results and image analyses, and in Section~\ref{section:discussion}, we discuss the observed pc-scale jet properties and their implications.  Finally, the summary is presented in Section~\ref{section:summary}.  Throughout this paper, a flat cosmology is assumed, with $H_0=71$~km~s$^{-1}$~Mpc$^{-1}$, $\Omega_\mathrm{M}=0.27$, and $\Omega_\mathrm{\Lambda}=0.73$ \citep{Spergel_etal.2003}.

\section{Sample}\label{section:sample}
\citet{Zhou&Wang2002} found nine radio-loud NLS1s from 205 NLS1s listed in ``A Catalogue of Quasars and Active Nuclei: 10th ed. \citep{Veron-Cetty&Veron2001}.''  Regarding these nine objects, we searched the Very Large Array~(VLA) archival data obtained at 4.9 and 8.4~GHz with the VLA in the A-array configuration, which could provide radio positions with sufficient accuracy ($\sim0\arcsec.1$) for processing in VLBI correlators.  We retrieved the data from the US National Radio Astronomy Observatory~(NRAO) and analyzed them.  As a result, we obtained radio positions for the five objects (Table~\ref{table1}), which are our samples for the VLBA observations (this paper) and the JVN observations \citepalias{Doi_etal.2007}.  Thus, our sample selection processes were VLBI-oriented.  No attempt was made to ensure sample completeness; a strict study on the general properties of the radio-loud NLS1s is out of the scope of this paper.   

The properties of optical emission lines that are used to classify the five objects as NLS1s in literatures are described in the Appendix; they are the FWHM of $\mathrm{H\beta}$ of the broad component, the flux ratio [\ion{O}{3}]/H$\beta$, and $R_\mathrm{4570}$, where $R_\mathrm{4570}$ is the flux ratio of the \ion{Fe}{2} multiplets in the range 4434--4684$\mathrm{\AA}$ \citep{Veron-Cetty_etal.2001}.

\begin{table}
\caption{Radio-loud NLS1 Samples for Our Observations\label{table1}}
\begin{center}
\begin{tabular}{lccccrc} \hline\hline
\multicolumn{1}{c}{Name} & $z$ & $F_\mathrm{20cm}$ & $F_\mathrm{6cm}$ & $\log{R_\mathrm{*}}$ & \multicolumn{1}{c}{$S_\mathrm{8.4GHz}^\mathrm{VLBI}$} & $\log{L_\mathrm{x}}$ \\
\multicolumn{1}{c}{} &  & (mJy) & (mJy) &  & \multicolumn{1}{c}{(mJy)} & (erg s$^{-1}$) \\
\multicolumn{1}{c}{(1)} & (2) & (3) & (4) & (5) & \multicolumn{1}{c}{(6)} & (7) \\\hline
RX~J0806.6+7248 & 0.099  & 50\tablenotemark{a}  & 26 & 2.07  & 5.0  & 43.9  \\
FBQS~J1629+4007 & 0.272  & 12  & 20 & 2.15  & 23.3  & 45.3  \\
RX~J1633.3+4718 & 0.116  & 65  & 35 & 1.87  & 22.1  & 44.3  \\
FBQS~J1644+2619 & 0.145  & 91  & 99 & 2.97\tablenotemark{b} & 150.3  & 44.1  \\
B3~1702+457 & 0.061  & 119  & 26 & 1.02  & 18.5  & 44.0  \\\hline
\end{tabular}
\end{center}
\tablecomments{Col.~(1) source name; Col.~(2) redshift; Col.~(3) 20-cm total radio flux density from the FIRST catalog \citep{White_etal.1997}; Col.~(4) 6-cm total radio flux density from the GB6 catalog \citep{Gregory_etal.1996}; Col.~(5) radio loudness, which was derived from the $k$-corrected 6-cm radio flux density and {\it V}-band magnitude using the values listed in \citet{Veron-Cetty&Veron2006} assuming an optical spectral index of $-0.5$; Col.~(6) VLBI flux density measured with the JVN at 8.4~GHz \citepalias[$\lambda 3.6$~cm; ][]{Doi_etal.2007}; Col.~(7) {\it ROSAT} X-ray luminosity at 0.1-2.4~keV \citep{Bade_etal.1995,Laurent-Muehleisen_etal.1998}.}
\tablenotemark{a}{Flux density at 1.4~GHz from the NRAO VLA Sky Survey~(NVSS; \citealt{Condon_etal.1998})}

\tablenotemark{b}{A much larger value compared to that in \citealt{Zhou&Wang2002}, mainly because they used $m_V=17.10$~mag from \citealt{Veron-Cetty&Veron2001}, while we used $m_V=18.44$~mag, on the basis of SDSS 4DR \citep{Adelman-McCarthy_etal.2006} from \citet{Veron-Cetty&Veron2006}.}
\end{table}

\section{OBSERVATIONS AND DATA REDUCTION}\label{section:observation}
The observations were performed using 10 antennas of the VLBA on 2006 February~07 and~10, while the JVN observations were conducted on 2006 March~17 and 26 and May~20 \citepalias{Doi_etal.2007}: the difference between the dates of the VLBA and JVN observations was $\sim$5--15~weeks.  A left circular polarization was obtained at a center frequency of 1.667~GHz with a total bandwidth of 32~MHz.  Because the targets were weak, we used a phase-referencing technique that involved fast switching of an antenna's pointing direction between a target and a reference calibrator in order to remove atmospheric fluctuation for increasing the coherent integration time.  The switching-cycle period was usually 5~minutes or $\sim$3~minutes at low elevations\footnote{To make phase-referencing secure, more than one calibrator was used for each target \citep{Fomalont&Kopeikin2003, Doi_etal.2006b} during the JVN observation.  However, this was not advantageous for the VLBA observation and significantly good results have been obtained without the secondary calibrators.}.      

Data reduction and imaging procedures were performed using the Astronomical Image Processing System~(AIPS; \citealt{Greisen2003}) and Difmap software \citep{Shepherd1997}, respectively.  Amplitude calibration using a priori gain values together with the system noise temperatures measured during the observations was applied, and the calibration accuracy was typically 5\%.  Inappropriate earth-orientation parameters in the correlator were corrected.  The geometrical correction of the antenna parallactic angles for the rotating feeds on the altitude-azimuth antennas in the position angle (P.A.) with respect to the source during the course of the observation was performed.  Ionospheric dispersive delay was corrected with a global ionospheric model using the task TECOR.  Careful data inspection in time and spectral domains and data flagging were performed; there was no significant radio interference in the data.  Bandpass correction was performed using the amplitude and phase data of a strong calibrator.  The fringe-fitting solution for the reference calibrator was obtained and then applied to the target~(``phase-referencing'').  We found that the phase-referenced visibilities of the targets had significantly high signal-to-noise ratios for the self-calibration of the integrations for $<$30~minutes.  We performed imaging with Difmap, which iteratively uses deconvolution and self-calibration algorithms for both amplitude and phase, starting from the phase-referenced data and a point source model.

\section{Results}\label{section:result}
We detected all five radio-loud NLS1s.  Images with significantly much higher dynamic ranges than those of the JVN were obtained.  The resultant images are shown in Figure~\ref{figure1}; the image parameters are listed in Table~\ref{table2}.  At 1.7~GHz, all targets showed some structures in the VLBA images (see the Appendix for details on individual objects), whereas they were featureless in the JVN images at 8.4~GHz~\citepalias{Doi_etal.2007}.  The results of the measurements of peak intensities and flux densities are listed in Table~\ref{table3}.  In this paper, we define peak intensity as the core in each image.

\subsection{Brightness Temperatures}\label{section:brightnesstemperature}
We calculated the brightness temperature for the core of each source.  The brightness temperature in K of a component at the source frame is given by 
\begin{equation}
T_\mathrm{B} = 1.8 \times 10^9 (1+z) \frac{S_\nu}{\nu^2 \phi_\mathrm{maj} \phi_\mathrm{min}},  
\label{equation:brightnesstemperature}
\end{equation}
where $z$ is the redshift, $S_\nu$ is the observed flux density in mJy at frequency $\nu$ in GHz, and $\phi_\mathrm{maj}$ and $\phi_\mathrm{min}$ in mas are the fitted FWHMs of the major and minor axes of the source size, respectively \citep{Ulvestad_etal.2005}.  Because the core in each source was unresolved, we adopted one-half of the beam sizes (i.e., $\theta/2$; Table~\ref{table2}) as the upper limits to the source sizes $\phi$.  The calculated brightness temperatures are listed in Table~\ref{table3}; $T_\mathrm{B} > 10^{8.7}$~K was derived for all sources.  

The geometrical mean of the synthesized beam sizes of the VLBA images was $4.6 \times 10.2$~mas and that of the image sensitivities was 0.144~mJy~beam$^{-1}$ (Table~\ref{table2}), which was equivalent to the brightness temperature $2.0 \times 10^6$~K.

\subsection{Spectral Indices}\label{section:spectralindices}
By comparing the flux densities (sum of the clean components), the spectral index~$\alpha$~($S_\nu \propto \nu^{+\alpha}$) can be obtained (Table~\ref{table3}).  The ranges of spatial frequency ($uv$-range), the ratios of the projected-baseline length to the wavelength, are $\sim1.0$--50~M$\lambda$ at 1.7~GHz for the VLBA observations and $\sim4.0$--65~M$\lambda$ for the JVN observations at 8.4~GHz, which corresponds to the scales of 4.1--210~mas and 3.2--52~mas, respectively.  The similar $uv$-ranges and the proximity in time (difference of only $\sim$5--15~weeks) of the VLBA and JVN observations can minimize the systematic errors related to the spatial resolution effects and flux variability in deriving a spectral index.  Only \object{RX J0806.6+7248} might have a possible structure larger than 52~mas, which may be resolved out in the JVN baselines at 8.4~GHz.  In addition, the lower sensitivity of the JVN images could derive the less clean components found above the image-noise levels of the images of diffuse emissions such as RX~J0806.6+7248.  Hence, the derived spectral indices indicate at least a lower limit for RX~J0806.6+7248.

\begin{table*}
\caption{Parameters of VLBA Images\label{table2}}
\begin{center}
\begin{tabular}{lccrcc} \hline\hline
\multicolumn{1}{c}{Name} & $\sigma$ & $\theta_\mathrm{min}\times\theta_\mathrm{maj}$ & \multicolumn{1}{c}{P.A.} & DR & $l$ \\
\multicolumn{1}{c}{} & (mJy beam$^{-1}$) & (mas$\times$mas) & \multicolumn{1}{c}{($\degr$)} &  & (pc mas$^{-1}$) \\
\multicolumn{1}{c}{(1)} & (2) & (3) & \multicolumn{1}{c}{(4)} & (5) & (6) \\\hline
RX~J0806.6+7248 & 0.117 & $4.7\times8.4$ & $-20.5$ & 82  & 1.8  \\
FBQS~J1629+4007 & 0.148 & $4.1\times10.7$ & $-10.1$ & 70  & 4.1  \\
RX~J1633.3+4718 & 0.114 & $4.8\times11.6$ & $35.3$ & 429  & 2.1  \\
FBQS~J1644+2619 & 0.176 & $4.5\times9.2$ & $-4.1$ & 446  & 2.5  \\
B3~1702+457 & 0.166 & $4.7\times11.4$ & $-47.5$ & 357  & 1.2  \\\hline
\end{tabular}
\end{center}
\tablecomments{Col.~(1)~source name; Col.~(2)~rms of image noise; Col.~(3)~FWHMs of minor and major axes of the synthesized beam; Col.~(4)~position angle of the beam major axis; Col.~(5)~image dynamic range, defined as the ratio of peak intensity (Table~\ref{table3}) to the rms of image noise; Col.~(6)~linear scale in pc corresponding to 1~mas at the distance to the source.}
\end{table*}

\begin{table*}
\caption{Results of VLBA Observations at 1.7~GHz\label{table3}}
\begin{center}
\begin{tabular}{lcccccccc} \hline\hline
\multicolumn{1}{c}{Source name} & $I^\mathrm{VLBI}_\mathrm{1.7GHz}$ & $S^\mathrm{VLBI}_\mathrm{1.7GHz}$ & $\log{P_\mathrm{1.7GHz}}$ & $\log{T_\mathrm{B}^\mathrm{core}}$ & $\alpha_I$ & $\alpha_S$ \\
\multicolumn{1}{c}{}  & (mJy beam$^{-1}$) & (mJy) & (W Hz$^{-1}$) & (K) &  &  \\
\multicolumn{1}{c}{(1)} & (2) & (3) & (4) & (5) & (6) & (7) \\\hline
RX~J0806.6+7248 & 9.6 $\pm$ 0.5  & 23.0 $\pm$ 1.2  & 23.7  & $>$8.7 & $-$0.51 $\pm$ 0.23 & $-$0.94 $\pm$ 0.42 \\
FBQS~J1629+4007 & 10.4 $\pm$ 0.5  & 10.5 $\pm$ 0.5  & 24.2  & $>$8.9 & $+$0.59 $\pm$ 0.19 & $+$0.49 $\pm$ 0.26 \\
RX~J1633.3+4718 & 48.9 $\pm$ 2.4  & 55.5 $\pm$ 2.8  & 24.3  & $>$9.3 & $-$0.73 $\pm$ 0.18 & $-$0.57 $\pm$ 0.20 \\
FBQS~J1644+2619 & 78.4 $\pm$ 3.9  & 100.6 $\pm$ 5.0  & 24.7  & $>$9.8 & $+$0.38 $\pm$ 0.16 & $+$0.25 $\pm$ 0.16 \\
B3~1702+457 & 59.2 $\pm$ 3.0  & 79.9 $\pm$ 4.0  & 23.8  & $>$9.4 & $-$0.84 $\pm$ 0.18 & $-$0.90 $\pm$ 0.21 \\\hline
\end{tabular}
\end{center}
\tablecomments{Col.~(1)~source name; Col.~(2)~peak intensity; Col.~(3)~flux density of total CLEANed components; Col.~(4) radio power at a rest frequency of 1.7~GHz, in which a {\it k}-correction was applied assuming the spectral index shown in column~(7); Col.~(5)~brightness temperature of the core component in the rest frame (Equation~(\ref{equation:brightnesstemperature})); Col.~(6)~spectral index derived from VLBA and JVN peak intensities; Col.~(7)~spectral index derived from VLBA and JVN flux densities.}
\end{table*}

To derive the spectral indices of the compact core components, we also express the spectral indices in terms of peak intensities, which defined as the strongest intensity, between 1.7~GHz and 8.4~GHz in the images.  The resolution effect is presumably negligible for these unresolved components.

\begin{figure*}
\epsscale{0.9}
\plotone{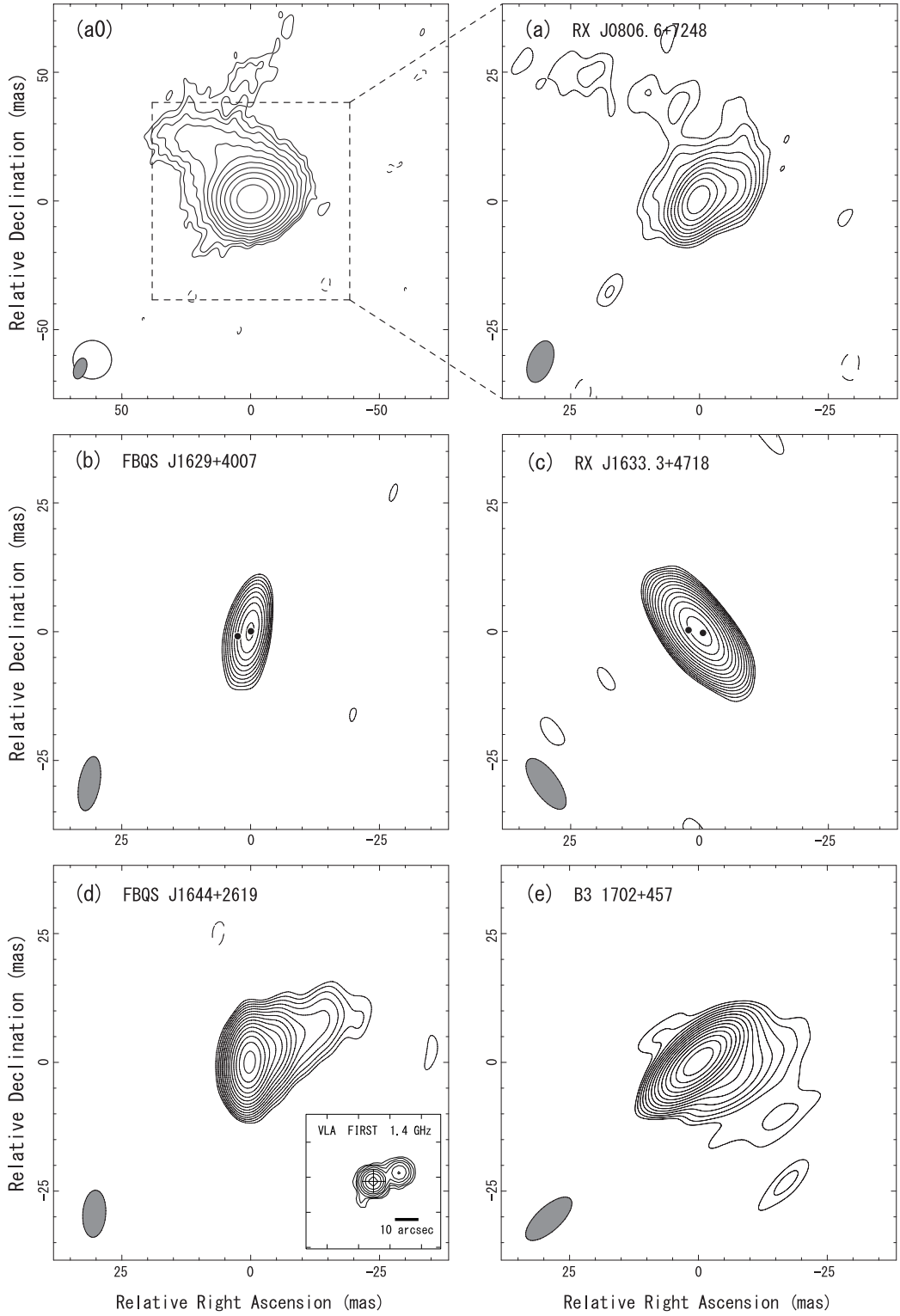}
\figcaption{VLBA images of five radio-loud NLS1s at 1.7~GHz.  Source names are indicated at the upper left corner in each panel.  All images were formed using natural weighting and were self-calibrated.  Contour levels are separated by factors of $\sqrt{2}$ beginning at three~times the rms~noise~(Table~\ref{table2}).  Negative and positive contours are shown as dashed and solid curves, respectively.  Half-power beam sizes are given in the lower left corners.  A large-scale map in panel~(a0) was created by convolution using a restored beam of 15~mas (shown as an open circle) for easy recognition of the low surface brightness structures.  The positions of the model-fitted components of FBQS~J1629+4007 and RX~J1633.3+4718 are indicated as points.\label{figure1}}
\end{figure*}

\section{DISCUSSION}\label{section:discussion}
We found significantly resolved pc-scale radio structures for three NLS1s and compact structures for two NLS1s.  The key results from our VLBA images are as follows:  (1)~detections of very high brightness temperatures and linear structures of the pc-scale radio emissions, which provide evidence of nonthermal jets in the NLS1s~(Section~\ref{section:nonthermaljets}); (2)~asymmetry of radio structures and core spectral indices that are very important information for evaluating the Doppler beaming effect on the jets (Section~\ref{section:asymmetricstructure}); and (3)~discovery of different radio properties of the five sources, which suggest why these NLS1s are radio loud~(Section~\ref{section:naturesofRLNLS1s}).  Detailed descriptions of individual objects are given in the Appendix.

\subsection{Strong Evidence for Nonthermal Jets}\label{section:nonthermaljets}
The previous JVN observations at 8.4~GHz detected brightness temperatures of at least $10^{7.4}$~K and high radio luminosities, which cannot be attributed to any stellar process; radio emissions presumably originate from nonthermal jets.  The VLBA observations have also revealed that radio components with very high brightness temperatures, which provide stronger constraints, i.e., $T_\mathrm{B}>10^{8.7}$~K, were associated with the five NLS1s.  Significant pc-scale radio structures from the VLBA observations were newly detected, providing stronger evidence for the presence of nonthermal jets.  \object{FBQS J1644+2619} and B3~1702+457 clearly displayed linear structures. (RX~J0806.6+7248 also showed a linear structure in the east and west of its nuclear region.)  

RX~J0806.6+7248, \object{RX J1633.3+4718}, and \object{B3 1702+457} are steep-spectrum sources~($\alpha < -0.5$) with very high brightness temperatures.  This is a strong evidence for a nonthermal process dominating the radio emissions.  On the other hand, \object{FBQS J1629+4007} and FBQS~J1644+2619 showed inverted spectra~($\alpha>0$).  The observed brightness temperatures were too high for free--free emissions, although a radio spectrum of not only nonthermal synchrotron emissions but also thermal free--free emissions could be inverted in an optically thick regime\footnote{For example, a thermal torus of at most $T_\mathrm{B} \sim 10^6$~K such as the one inferred for type-2 Seyfert NGC~1068 \citep{Gallimore_etal.1997,Gallimore_etal.2004} was imaged using the VLBA.}.  We consider a putative source with having a diameter of lower than that of the VLBA beams radiating the free--free emissions at electron temperatures equal to the observed brightness temperatures and being optically thick at our observing frequency.  Under these conditions, the X-ray luminosities at 0.1--2.4~keV of the bremsstrahlung emission would be $\sim10^{48}$~erg~s$^{-1}$, which cannot meet the observed X-ray luminosities $\sim10^{44}$~erg~s$^{-1}$ (Table~\ref{table1}); these were detected in the {\it ROSAT} All Sky Survey \citep{Bade_etal.1995,Laurent-Muehleisen_etal.1998} and should include the contribution not only of bremsstrahlung emission but also the inverse Compton and synchrotron emissions.  Hence, for all observed NLS1s, we found strong evidence that the nonthermal process dominates the observed radio emissions.  

Because the VLBI-correlated fluxes at 1.7~GHz were equivalent to much of the total fluxes in the VLA at 1.4~GHz, we conclude that the radio emissions of all these radio-loud NLS1s are dominated by the nonthermal synchrotron process in jets and that these pc-scale jets make the NLS1s radio loud.

\subsection{Asymmetric Core--Jet Structures}\label{section:asymmetricstructure}
The AGN radio sources generally consist of steep-spectrum jets ($\alpha<0$) and a flat-/inverted-spectrum core ($\alpha \ga 0$); the core is the location where the synchrotron emission becomes visible from an optically thick upstream to an optically thin downstream at an observing frequency.  The core flux contribution generally tends to a relative increase in the higher frequency radio images; therefore, the JVN emissions should originate from cores.   

A structural asymmetry relative to the core was found in the limited dynamic ranges of our images, although only two components were seen in FBQS~J1629+4007 and RX~J1633.3+4718.  The possible causes of the asymmetry are as follows: (1)~the jets are intrinsically one-sided or more lossy (and hence radiative) on one side than the other; or (2)~Doppler boosting effect on the relativistic jets; or (3)~the jets are intrinsically symmetric but appear asymmetric owing to the stronger free-free absorption toward the counter-jets due to an obscuring ionized gas intervening along our line of sight \citep{Walker_etal.2000,Jones et al.2001,Kameno_etal.2001}.  Both (2) and (3) require that the stronger side should be approaching jets, while the weaker side should be receding jets.  In this paper, we discuss the possible parameters of the jets in the NLS1s on the assumption that the jet are intrinsically symmetric.  The images of RX~J0806.6+7248 and B3~1702+457 showed possible counter features, and thus indicating receding jets.  However, it should be noted that in some cases, the VLBI imaging processes probably generate a mimic counter component near the core \citep[e.g., imaging demonstrations by][]{Carilli_etal.1994}.  We cannot be confident that the counter-jet features were genuine receding jets (the counter jets in RX~J0806.6+7248 may exist because a diffuse counter feature, which is misaligned with the linear structure, was also seen at a small distance; the Appendix.  On the other hand, no counter component can be seen in FBQS~J1644+2619, which has a significant linear jet structure in the limited image dynamic range; this result provides a reliable constraint for a receding jet.  Hereafter, we consider FBQS~J1644+2619 for discussing asymmetry.

\subsubsection{Doppler Beaming Effect Estimated from One-Sidedness}\label{section:Dopplerbeamingeffect}
In case Doppler boosting was responsible for the observed asymmetry of the jet structure, we can limit the jet speed and inclination by comparing the approaching and receding jets on the assumption that the rest-frame radio spectra and intrinsic jet speeds of both the sides were the same.  A bulk Lorentz factor is defined as $\Gamma \equiv (1-\beta^2)^{-0.5}$, where $\beta \equiv v/c$ ($v$ and $c$ are the speeds of the source and light, respectively).  The Doppler factor is defined as 
\begin{equation}
\delta \equiv \frac{1}{\Gamma (1-\beta\cos{\Phi})}, 
\end{equation}
where $\Phi$ is the jet viewing angle, which is the angle between the direction of the source velocity and our line of sight \citep[][Appendix~B]{Ghisellini_etal.1993}.  The frequency and the flux density at the rest frame, $\nu_\mathrm{rest}$ and $S_{\nu\mathrm{rest}}^\mathrm{rest}$, are observed as those at the observer frame, $\nu_\mathrm{obs} = \delta \nu_\mathrm{rest}$ and $S_{\nu\mathrm{obs}}^\mathrm{obs} = \delta^3 S_{\nu\mathrm{rest}}^\mathrm{rest}$, respectively.  Assuming the spectral index $\alpha$, $S_{\nu\mathrm{obs}}^\mathrm{obs} = \delta^{3-\alpha} S_{\nu\mathrm{obs}}^\mathrm{rest}$ \citep{Rybicki&Lightman1979}.  Hence, the flux-density ratio of the approaching and receding jets is 
\begin{eqnarray}
r &=& \frac{S_{\nu\mathrm{obs}}^\mathrm{rest}(\beta, \Phi)}{S_{\nu\mathrm{obs}}^\mathrm{rest}(\beta, \Phi + \pi)}\nonumber\\
               &=& \left( \frac{1+\beta \cos{\Phi}}{1-\beta \cos{\Phi}} \right)^q,
\end{eqnarray}
where $q=3-\alpha$; when considering the brightness ratio instead of the flux-density ratio in continuous jets $q=2-\alpha$ is applied \citep{Ghisellini_etal.1993}.  Therefore, we obtained 
\begin{equation}
\beta \cos{\Phi} > \frac{r^{1/q}-1}{r^{1/q}+1}.\label{eq:fluxratio}
\end{equation}
Using this inequality, we can estimate the lower limit of $\beta$ and the upper limit of $\Phi$.

\begin{table}
\caption{Parameters of the Doppler Beaming Effect for FBQS~J1644+2619\label{table4}}
\begin{center}
\begin{tabular}{ccccccc} \hline\hline
Source name & Comparison & $r$ & $k$ & $\alpha$ & $\beta_\mathrm{min}$ & $\Phi_\mathrm{max}$ \\
 &  &  &  &  &  & ($\degr$) \\
(1) & (2) & (3) & (4) & (5) & (6) & (7) \\\hline
\multicolumn{1}{l}{FBQS~J1644+2619} & $S_{\nu}(\mathrm{core})/3\sigma$  & $>$147 & 3  & $\alpha_I$ & 0.74  & 42  \\
$\uparrow$ & $I_{\nu}(\mathrm{jet})/3\sigma$  & $>$37 & 2  & $\alpha_S$ & 0.77  & 39  \\\hline
\end{tabular}
\end{center}
\tablecomments{Col.~(1)~source name; Col.(2)~measurements compared to derive the ratio of jet to counter jet.  $S_{\nu}(\mathrm{core})$ means the core's flux density.  $3\sigma$ means three times the rms~image noise as an upper limit for a counter jet.  $I_{\nu}(\mathrm{jet})$ means the intensity of the jet emission at the position separated from the peak by the spatial resolution along the jets; Col.(3)~ratio of jet to counter jet; Col.~(4)~$k$ value used for the calculation of the Doppler beaming effect, where $q \equiv k- \alpha$ (Section~\ref{section:Dopplerbeamingeffect}); Col.~(5)~assumed spectral index (Table~\ref{table3}); Col.~(6) lower limit of jet speed (Equation~(\ref{eq:fluxratio})); Col.~(7)~upper limit of the angle of jet axis (Equation~(\ref{eq:fluxratio})).}
\end{table}

\begin{table*}
\caption{Parameters of the Variability Doppler-beaming Effect for FBQS~J1629+4007\label{table5}}
\begin{center}
\begin{tabular}{ccccccccccccc} \hline\hline
Name & $I^\mathrm{FIRST}_\mathrm{1.4}$ & Obs.~Date & $S^\mathrm{NVSS}_\mathrm{1.4}$ & Obs.~Date & Variability & $\log{T_\mathrm{B}}$ & $\log{T_\mathrm{int}^\mathrm{max}}$ & $\delta_\mathrm{min}$ & $\Phi_\mathrm{max}$ & $\beta_\mathrm{min}$ & $\log{P_\mathrm{1.7GHz}^\mathrm{int}}$ & $\log{R_\mathrm{*}^\mathrm{int}}$ \\
 & (mJy beam$^{-1}$) &  & (mJy) &  &  & (K) & (K) &  & ($\degr$) &  & (W Hz$^{-1}$) &  \\
(1) & (2) & (3) & (4) & (5) & (6) & (7) & (8) & (9) & (10) & (11) & (12) & (13) \\\hline
\multicolumn{1}{l}{FBQS~J1629+4007} & 11.97 $\pm$ 0.17 & 1994 Aug 21 & 9.0 $\pm$ 0.5 & 1995 Apr 16 & $5.6\sigma$ & 13.0  & 11.2  & 4.0  & 14  & 0.88  & $<$22.7 & $<$0.74 \\
$\uparrow$ & $\uparrow$ & $\uparrow$ & $\uparrow$ & $\uparrow$ & $\uparrow$ & $\uparrow$ & 12 & 2.1  & 28  & 0.63  & $<$23.4 & $<$1.43 \\\hline
\end{tabular}
\end{center}
\tablecomments{Col.~(1)~source name; Col.(2)~1.4-GHz peak flux density from FIRST; Col.~(3)~observation date of the FIRST data; Col.~(4)~1.4-GHz peak flux density from NVSS; Col.~(5)~observation date of the NVSS data; Col.~(6)~significance of the detection of variability; Col.~(7)~brightness temperature derived from variability; Col.~(8)~assumption of maximum intrinsic brightness temperature (Section~\ref{section:fluxvariability}); Col.~(9)~lower limit of Doppler factor; Col.~(10)~upper limit of viewing angle of jet axis; Col.~(11)~lower limit of jet speed; Col.~(12)~intrinsic radio power; Col.~(13)~intrinsic radio loudness.}
\end{table*}

Using the inequality, we derived the parameters of the Doppler effect for FBQS~J1644+2619 (Table~\ref{table4}) without considering the free--free absorption.  We considered that the core was the first visible component in the approaching jet stream and that the flux density of the counter component in the receding stream was less than three times the rms~image noise (Table~\ref{table2}).  We found that $\beta>0.74$, which is either a mildly or highly relativistic jet speed and that $\Phi<42$\arcdeg, which is a mild constraint for the inclination in this NLS1 (Equation~(\ref{eq:fluxratio})).  For the case of the brightness ratio in continuous jets ($q=2-\alpha$), we found a similar result of $\beta>0.77$ and $\Phi<39$\arcdeg.  This result indicates that the central engine of an NLS1 can generate nonthermal jets with quite a high speed.  The radio loudness of this NLS1 is potentially due to the Doppler beaming effect.  

We cannot obtain $\delta$ by only considering the apparent jet structure.  VLBI image monitoring will provide the apparent jet speed in order to solve both $\beta$ and $\Phi$ for obtaining $\delta$.  Another method to obtain the constraint $\delta$ by imaging is to confirm the compactness of the source at a higher spatial resolution.  For example, by VLBI observations, the very radio-loud ($R \approx 2000$) NLS1 SDSS~J094857.31+002225.4 has highly relativistic jets viewed pole-on because of its compactness, which is indicative of an extremely high brightness temperature requiring Doppler boosting of $\delta>$2.7--5.5 \citepalias{Doi_etal.2006a}.

\subsubsection{Free--Free Absorption}\label{section:freefreeabsorption}
The possibility of the apparent jet asymmetry because of the free-free absorption due to the intervening ionized thermal plasma is discussed for FBQS~J1644+2619.  The optical depth of free--free absorption is 
\begin{equation}
\tau_\mathrm{ff} \approx 8.235 \times 10^{-2} T_\mathrm{e}^{-1.35} \nu^{-2.1} \int n_\mathrm{e} dl,  
\end{equation}
where $T_\mathrm{e}$~(K) and $n_\mathrm{e}$~(cm$^{-3}$) are the electron temperature and density of the absorbing gas, respectively, $\nu$ is the frequency~(GHz), and $l$ is the path length in pc~\citep{Mezger_Henderson1967}.  The attenuation of more than a factor of 37 (Table~\ref{table4}) that corresponds to $\tau_\mathrm{ff}>3.6$ is required to make a putative counter-jet opaque.  As a rough estimate, considering a path length equivalent to the length of the putative counter-jet of 20~mas~(50~pc), the fully ionized gas with $T_\mathrm{e}>8000$~K, and $\nu=(1+z)\times1.7$~GHz, a large amount of electrons of the column density of electrons $N_\mathrm{e}(=n_\mathrm{e}l) > 10^{23}$~cm$^{-2}$ would be required in the nuclear region.  {\it Chandra} X-ray observations of FBQS~J1644+2619 showed that the fitted absorption column density $N_\mathrm{H}$ (natural gas) is almost identical to the Galactic value $5.12 \times 10^{20}$~cm$^{-2}$ \citep{Yuan_etal.2008}, suggesting a very poor intervening medium toward the nucleus.  Thus, such a large free--free absorber is unlikely in the FBQS~J1644+2619 nucleus unless an extremely ionized fraction is assumed.  Furthermore, the VLA image of FBQS~J1644+2619 (Figure~\ref{figure1}(d)) also seems to show an asymmetric structure, which is presumably related to the AGN activity~(the Appendix).  The large-scale asymmetry up to 30~kpc implies that the free-free absorption cannot be responsible for the apparent asymmetry, because such optically thick plasma would provide thermal radio emission of $\sim4$~Jy (Equation~(\ref{equation:brightnesstemperature})), which is much larger than the observed total flux density.  Hence, we conclude that Doppler beaming should be preferentially efficient and it contributes to making FBQS~J1644+2619 one-sided, unless the jets are intrinsically one-sided.  The intriguing large structure of FBQS~J1644+2619 is discussed later in Section~\ref{section:naturesofRLNLS1s}.

\subsection{Doppler Beaming Inferred from Flux Variability}\label{section:fluxvariability}
The Doppler factor can also be restricted by brightness temperature, which is derived from flux variability.  For our sample, we checked the flux variation between the NRAO VLA Sky Survey~(NVSS; \citealt{Condon_etal.1998}) and the Faint Images of the Radio Sky at Twenty centimeters~(FIRST; \citealt{Becker_etal.1995}) at 1.4~GHz.  Because of the resolution effect due to the smaller beam size of FIRST, we could find an intrinsic variability only if the flux density in FIRST was larger than that in NVSS.  Data for all sources except for RX~J0806.6+7248 were available from both NVSS and FIRST.  Only FBQS~J1629+4007 showed a FIRST flux density larger than that in NVSS.  Using the same technique as \citet{Ghosh&Punsly2007}, we detected a flux variation in FBQS~J1629+4007 with a significance of $5.6\sigma$; the brightness temperatures were derived to be $10^{13.0}$~K~(Table~\ref{table5}).  Based on the near equipartition of energies between radiating particles and a magnetic field \citep{Readhead1994} or the catastrophic electron energy loss due to an inverse Compton process \citep{Kellermann_Pauliny-Toth1969}, the brightness temperature of nonthermal radio emission is expected to be within a physically realistic upper limit of $10^{11.2}$~K or $10^{12}$~K in the rest frame; the Doppler factor is expected to be $\delta>4.0$ or 2.1 and the viewing angles should be 14$\arcdeg$ or 28$\arcdeg$, respectively.  Thus, it is evident that the emitting source is a highly relativistic jet viewed pole-on.  This source is presumably Doppler boosted and may be intrinsically in the radio-quiet population~(Column~13 of Table~\ref{table5}).  The variability Doppler factors for several objects have also been previously reported in a very radio-loud NLS1 population \citep{Yuan_etal.2008}.  Our limited dynamic-range VLBA image of FBQS~J1629+4007 showed only a structure of two discrete components (the Appendix), which probably indicated a one-sided jet.  It is important to conduct VLBI imaging more thoroughly and at more observing frequencies in the future.

\subsection{Nature of Radio-loud NLS1 Radio Sources}\label{section:naturesofRLNLS1s}
NLS1s are expected to be radio-quiet AGNs on the basis of the possible dependence of radio loudness on both the black hole mass and the accretion rate~\citep{Boroson&Green1992}.  Even if all NLS1s were intrinsically radio quiet, the Doppler beaming effect could explain why some NLS1s appeared radio loud.  Our VLBA study for the five radio-loud NLS1s has shown both the cases that are likely and unlikely due to the Doppler beaming effect.  Although they may be intrinsically radio quiet, FBQS~J1629+4007 and FBQS~J1644+2619 (inverted-spectrum sources) have shown evidence of a significant Doppler beaming effect, which presumably contributes to their radio loudness.  We emphasize that the strong Doppler beaming requires not only a nearly pole-on viewed orientation but also a relativistic bulk jet speed.  The nuclei of these NLS1s are expected to have the ability to accelerate the jets to the relativistic speed.  In the projected distance, the asymmetry of FBQS~J1644+2619 continues up to 30~kpc; kpc-scale asymmetry has also been seen in many blazars in their VLA images \citep{Murphy_etal.1993,Kharb_etal.2010}.  The jet length 30~kpc corresponds to $\sim 4 \times 10^{10}$~the Schwarzschild radius for FBQS~J1644+2619 (an estimated black hole mass of $8.4 \times 10^6\ M_\mathrm{\sun}$; \citealt{Yuan_etal.2008}).  The largest radio galaxies have radio structures of $>1$~Mpc \citep{Palma_etal.2000,Machalski_etal.2008}, which corresponds to $>1 \times 10^{10}$~the Schwarzschild radius of a black hole mass of $10^9\ M_\mathrm{\sun}$.  The significance of FBQS~J1644+2619 is that at least some NLS1 nuclei have the ability to be huge radio galaxies in black-hole-mass-normalized scales.  A further study of the kpc-scale jets of FBQS~J1644+2619 will be reported elsewhere.   

RX~J0806.6+7248 and B3~1702+457, which are steep-spectrum sources with non-negligible diffuse structures, are unlikely to be strongly beamed; this indicates intrinsic radio-loud sources without Doppler boosting.  RX~J0806.6+7248 shows jet bending at $\sim$20~pc from the core; this jet bending extends to $\sim$100~pc~(Figure~\ref{figure1}).  Such a structure resembles that of TeV gamma-ray blazar Mrk~501, which has inner jets strongly bending at $\sim30$~mas from the core and continuing onto very diffuse outer jets.  \citet{Giroletti_etal.2004} suggested that highly relativistic bulk speeds and a progressive change in the jet direction ($4$--$25\degr$) with respect to the line of sight are possible and in agreement with the observed one-sided jet properties in the pc scale of Mrk~501.  A quite small viewing angle is required to make such an intrinsically small bending appear large.  The core spectrum inverted up to 8.4~GHz \citep{Giroletti_etal.2004} can attribute to a strong beaming effect on the spectral peak frequency $\nu_\mathrm{obs} = \delta \nu_\mathrm{rest}$ (Section~\ref{section:Dopplerbeamingeffect}) for pole-on viewed innermost jets.  On the other hand, the inner jets of RX~J0806.6+7248 appear two-sided with respect to the core with a steep spectrum, and the extended jets originate from the both sides of the two-sided structure (the Appendix); the extended jets of Mrk~501 distinctly originate from the one-sized inner jets.  It is difficult to attribute these radio properties of RX~J0806.6+7248 to beamed jets.

Thus, the nuclei of some NLS1s can generate jets that are strong enough to make them intrinsically radio loud.  Consequently, either case (strongly or weakly beamed) suggests significant jet productivity in the radio-loud NLS1s in our sample.  NLS1s as a class show a number of extreme properties and radio-loud ones are very rare.  However, we build on these radio results to understand that the central engines of radio-loud NLS1s are essentially the same as that of other radio-loud AGNs in terms of the formation of nonthermal jets.

NLS1s show two types of X-ray continuum spectra.  Normal~(radio-quiet) NLS1s show a prominent soft excess plus a relatively steep ($\Gamma \sim 2$) power-law hard tail \citep{Haba_etal.2008,Leighly1999b}.  Both the components are proposed to originate from the accretion disks; the soft excess is thermal emission from the extremely slim disk \citep{Abramowicz_etal.1988,Mineshige_etal.2000,Wang&Netzer2003} and the power-law hard component originates from the scattered-up disk photons generated by the inverse Compton process.  On the other hand, \citet{Yuan_etal.2008} recently indicated that while some radio-loud NLS1s show X-ray spectra similar to normal NLS1s, many radio-loud NLS1s resemble the high-energy-peaked flat-spectrum radio quasars~\citep[HFSRQs;][]{Perlman_etal.1998} in terms of the broadband spectral energy distribution~(SED), which shows synchrotron spectra (due to beamed jets) peaked at UV/X-ray regimes, resulting in very steep X-ray spectra.  FBQS J1629+4007 and FBQS~J1644+2619, both of which are presumably Doppler beamed sources in our study, show HFSRQ-like SEDs \citep{Padovani_etal.2002,Yuan_etal.2008}.  In particular, FBQS~J1629+4007 is the first confirmed HFSRQ for its modeled synchrotron peak at $2\times10^{16}$~Hz and its very steep X-ray spectrum.  \citet{Zhou_etal.2007} also suggested that the radio-loud NLS1 J0324+3410 was an NLS1-blazar composite with an HFSRQ-like SED.  Although a small number of sources in our study are discussed here, the radio-loud NLS1s with beamed jets may have jet-dominated HFSRQ-like SEDs.  

On the other hand, RX~J1633.3+4718, which may be unlikely to be strongly Doppler beamed because of its steep spectrum, shows an X-ray spectrum with a soft excess plus a power-law hard tail similar to several normal NLS1s \citep{Yuan_etal.2008}.  \citet{Yuan_etal.2010} showed this ultra-steep soft X-ray emission can be well fitted by an optically thick accretion disk, whose inferred parameters (a black hole mass of $\sim 3 \times 10^6\ M_\mathrm{\sun}$ and accretion rate) are in good agreement with the independent estimates based on an optical emission line spectrum.  \citet{Yuan_etal.2008} discussed that the hard tail is flat~($\Gamma=1.37 \pm 0.49$) and may be an inverse Compton component originating from the jets, similar to the flat-spectrum radio quasars.  \citet{Abdo_etal.2009a} also discussed that the simultaneous optical/X-ray/$\gamma$-ray SED of the very radio-loud NLS1 SDSS~J094857.31+002225.4 showed a blue continuum attributed to the accretion disk and a hard X-ray spectrum attributed to the jet.  Thus, the contribution of the disk could not be negligible even in the radio-loud NLS1s.  Perhaps, even the HFSRQ-like SED may be due to only the dominance of the thermal emission from the extremely slim disk.  We would have to compare many SEDs and beamed-jet profiles by conducting more VLBI imaging observations for radio-loud NLS1s in order to understand the origin of the X-ray spectra.  

Similar to NLS1s, broad absorption line~(BAL) quasars \citep{Weymann_etal.1991} are also considered to be highly accreting black hole (with larger masses compared to NLS1s) systems \citep{Boroson2002} and generally radio quiet \citep{Becker_etal.2001}.  The rarity of the BAL quasars as radio galaxies \citep{Gregg_etal.2006} is an important similarity to NLS1s.  A VLBI study of BAL quasars \citep{Jiang_Wang2003,Montenegro-Montes_etal.2009,Doi_etal.2009,Kunert-Bajraszewska_etal.2010} is crucial for revealing their accreting and outflowing phenomena and determining their speed and orientation of the nonthermal jets by observing them.  An understanding of the natures of the highly accreting systems is required to explain both the subclasses of NLS1s and BAL quasars.

\section{Summary}\label{section:summary}
The VLBA observations at 1.7~GHz by direct imaging of pc-scale radio structure have provided convincing evidence of nonthermal jets in radio-loud NLS1s.  We resolved significant pc-scale structures of three NLS1s; the pc-scale radio emissions provided most of the total fluxes found in the VLA resolutions in all the targets.  The previously reported JVN images at 8.4~GHz enabled us to determine that our sample contained two inverted- and three steep-spectrum sources.  FBQS~J1644+2619, an inverted spectrum source, showed a remarkable one-sided pc-scale jet structure that indicated either a mildly or highly relativistic jet speed of $\beta>0.74$ and an inclination of $<42\degr$ on the basis of the framework of the Doppler beaming effect.  FBQS~J1629+4007, an inverted-spectrum source, showed compact radio structures in the pc-scale and a significant flux variability between the data from NVSS and FIRST.  The physically unrealistic brightness temperature inferred from the variability requires a pole-on viewed relativistic jet of a Doppler factor of $\delta>4.0$, implying a jet speed of $\beta>0.88$ and an inclination of $<14\degr$.  The Doppler beaming effect is possibly responsible for the radio loudness in FBQS~J1644+2619 and FBQS~J1629+4007.  RX~J0806.6+7248 and B3~1702+457, which are steep-spectrum sources with non-negligible diffuse structures in the pc scale, were unlikely to be strongly beamed; this indicates that they are intrinsically radio loud even without the beaming effect.  We found both types of radio-loud NLS1s with relativistically beamed jets and intrinsically strong jets, as well as seen in other radio-loud AGN classes.

\acknowledgments

We used of the NASA's Astrophysics Data System Abstract Service and the NASA/IPAC Extragalactic Database, which is operated by the Jet Propulsion Laboratory; we have also used Ned Wright's online cosmology calculator.  The VLBA and VLA are operated by the National Radio Astronomy Observatory, which is a facility of the National Science Foundation operated under cooperative agreement by Associated Universities, Inc.  The JVN project is led by the National Astronomical Observatory of Japan, which is a branch of the National Institutes of Natural Sciences, Hokkaido University, Gifu University, Yamaguchi University, Kagoshima University, the University of Tsukuba, Ibaraki University, and Osaka Prefecture University, in cooperation with the Geographical Survey Institute, the Japan Aerospace Exploration Agency, and the National Institute of Information and Communications Technology.

\appendix

\section{Individual Objects}\label{section:individualobjects}

We describe below the optical spectroscopic classification in the literature, the radio properties observed by our observations in terms of the pc-scale structure, and the radio spectrum of each NLS1, particularly from the viewpoint of the Doppler beaming effect.  

\subsection{RX J0806.6+7248} 
The optical emission line properties of RX~J0806.6+7248 are FWHM(H$\beta$)=1060~km~s$^{-1}$, [\ion{O}{3}]/H$\beta<3$, the unambiguously presence of strong \ion{Fe}{2} lines; thus, it is classified as an NLS1 \citep{Xu_etal.1999}.  Our VLBA image of RX~J0806.6+7248 at 1.7~GHz showed the most remarkable radio structure in our sample.  Low-brightness emissions with temperatures $10^6$--$10^7$~K extended northeast up to 100~pc from the intensity peak (Figure~\ref{figure1}(a0)).  The inner region of this source (Figure~\ref{figure1}(a)) displayed an east--west linear structure that consisted of four discrete components with high brightness temperatures ($>10^8$~K), which were identified by visibility-based model fitting.  We cannot completely rule out the possibility of a continuous linear structure.  Such structures in this steep-spectrum source (Table~\ref{table3}) are direct evidence that nonthermal jets are responsible for the observed radio emission and for making this NLS1 radio loud (a substantial fraction of the total flux was not resolved out in the VLBA image; Tables~\ref{table1} and \ref{table3}).  A two-sided structure with respect to the core can be seen in the linear structure.  We note that diffuse emissions appear to emanate toward the north from both ends of the inner linear structure, which supports a genuine two-sidedness of the inner linear structure.  The steep spectrum core and the two-sidedness are unlikely to be caused by a strong Doppler beaming effect in this NLS1, i.e., the jets are sub-relativistic or significantly inclined.

\subsection{FBQS J1629+4007} 
The optical emission line properties of FBQS~J1629+4007 are FWHM(H$\beta)=1238\pm20$~km~s$^{-1}$, [\ion{O}{3}]/H$\beta=439\pm5/1563\pm17$, and $R_\mathrm{4570}=0.48\pm0.02$, which were analyzed on the basis of Sloan Digital Sky Survey (SDSS) DR3; thus, it is classified as an NLS1 (\citealt{Zhou_etal.2006}; see also \citealt{Whalen_etal.2006}).  Our VLBA image at 1.7~GHz showed an apparently compact pc-scale structure, but 
visibility-based model fitting indicated two discrete components ($\sim10$~mJy and $\sim1$~mJy) rather than a single elliptical Gaussian (Figure.~\ref{figure1}(b)).  The weaker component was seen at 6.2$\sigma$ in a residual image after the subtraction of the stronger component, and was separated at $\mathrm{P.A.}=110\pm16\degr$ by $2.7\pm0.6$~mas~(11~pc), which is more than half of the synthesized beam.  A VLBA image at 5~GHz also shows a slightly resolved component separated by 1.5~mas at $\mathrm{P.A.}=122\degr$ \citep{Gu&Chen2010}.  The consistent position angles may imply the jet direction of this NLS1.  Because the total flux was detected in the VLBA image (Tables~\ref{table1} and \ref{table3}), the radio emission of this NLS1 was dominated by these pc-scale components with the high brightness temperatures~($>10^8$~K).  Although the jet feature was insufficient to discuss one/two-sidedness, the combination of the inverted radio spectrum and the very compact structure was suggestive of the Doppler beaming effect.  We also show further evidence of Doppler beaming with the Doppler factor $\delta>4.0$ (jet speed $>0.88c$) from flux variability in the NLS1 in Section~\ref{section:fluxvariability}.

\subsection{RX J1633.3+4718} 
The optical emission line properties of RX~J1633.3+4718 are FWHM(H$\beta)=909\pm43$~km~s$^{-1}$, [\ion{O}{3}]/H$\beta=902\pm21/918\pm8$, $R_\mathrm{4570}=1.02$, which were analyzed on the basis of SDSS DR5; thus, it is classified as an NLS1 (\citealt{Yuan_etal.2008}; see also \citealt{Wisotzki&Bade1997, Zhou_etal.2006}).  Our VLBA image of RX~J1633.3+4718 at 1.7~GHz showed an apparently compact pc-scale structure, but indicated two discrete components ($\sim38$~mJy and $\sim18$~mJy) rather than a single elliptical Gaussian in the visibility-based model fitting (Figure~\ref{figure1}(c)).  The weaker component was seen at 11$\sigma$ in a residual image after the subtraction of the stronger component and was separated at $\mathrm{P.A.}=79\degr\pm2\degr$ by $2.9\pm0.1$~mas~(6.1~pc), about half of the synthesized beam.  A VLBA image at 5~GHz also showed a second component separated by 3.7~mas at $\mathrm{P.A.}=79\degr$ from the core \citep{Gu&Chen2010}.  The consistent position angles may imply the jet direction of this NLS1.  Because most of the total flux was detected in the VLBA image (Tables~\ref{table1} and \ref{table3}), the radio emission of this NLS1 was dominated by the pc-scale structure with very high brightness temperatures~($>10^9$~K).  Unlike FBQS~J1629+4007, this NLS1 showed a steep spectrum, despite the very compact radio structure.  Such radio properties cannot generally be attributed to strong Doppler beaming but were associated with the compact radio lobes with a sub-relativistic speed similar to that seen in the compact steep-spectrum objects, which are probably an AGN class of young radio galaxies~\citep{Odea1998}.

\subsection{FBQS J1644+2619} 
The optical emission line properties of FBQS~J1644+2619 are FWHM(H$\beta)=1507\pm42$~km~s$^{-1}$, [\ion{O}{3}]/H$\beta=120\pm4/1108\pm22$, and $R_\mathrm{4570}=0.75$, which were analyzed on the basis of SDSS DR5; thus, it is classified as an NLS1 (\citealt{Yuan_etal.2008,Whalen_etal.2006}).  Our VLBA image of FBQS~J1644+2619 at 1.7~GHz showed a continuous linear pc-scale structure with a length of $\sim20$~mas~($\sim50$~pc; Figure~\ref{figure1}(d)); the discrete components cannot be identified, and the width in the direction perpendicular to the linear structure cannot be resolved.  No diffuse emission component was found even in a fairly high dynamic range~($\mathrm{DR}=429$; Table~\ref{table2}); on the other hand, RX~J0806.6+7248 showed significant diffuse emissions even with a much lower dynamic range~($\mathrm{DR}=82$; Table~\ref{table2}).  The clear linear structure with high brightness temperatures ($>10^8$~K) was a direct evidence that nonthermal jets were responsible for the observed radio emission in this NLS1 and for making it radio loud (most of the total flux was detected in the VLBA image; Tables~\ref{table1} and \ref{table3}).  The one-sidedness together with the inverted spectrum was highly suggestive of the Doppler beaming effect.  We determined the jet speed $>0.74c$ (Section~\ref{section:Dopplerbeamingeffect}).  The one-sidedness seems to be maintained up to the kpc scale, shown in the FIRST archival image (Figure~\ref{figure1}(d)); this image was also shown by \citet{Whalen_etal.2006}.  Using the AIPS task JMFIT, we identified two radio components separated by 11.7~arcsec, corresponding to $\sim$30~kpc.  The brighter component was almost unresolved and coincident with the position of the VLBI emissions (a cross symbol in the FIRST image), while the secondary component was slightly resolved.  There is no optical counterpart for the secondary component in the NASA/IPAC Extragalactic Database.  The P.A.~(71\arcdeg) and the structural elongation~(72\arcdeg) of the secondary component are consistent with the direction of the pc-scale structure (61\arcdeg.0--63\arcdeg.6).  Hence, the secondary component was likely to physically relate to the nucleus of FBQS~J1644+2619.  

\subsection{B3 1702+457} 
The optical emission line properties of B3~1702+457 are FWHM(H$\beta)=1040$~km~s$^{-1}$, [\ion{O}{3}]/H$\beta=2.48$, and $R_\mathrm{4570}=1.86$; thus, it is classified as an NLS1 (\citealt{Leighly1999b,Veron-Cetty_etal.2001, Wisotzki&Bade1997}).  Our VLBA image of B3~1702+457 showed an elongated structure in the southwest direction with a length of $\sim30$~mas~($\sim35$~pc; Figure~\ref{figure1}(e)); the elongated structure seemed to curve from the west to the south.  The VLBA images at 5~GHz reported by \citet{Gu&Chen2010} showed a second weak component separated toward the west from an intensity peak in a much higher spatial resolution, which was consistent with our 1.7~GHz image around the core of the curved jets.  Unlike FBQS~J1644+2619 with a linear structure, the radio structure consisted of quite diffuse components ($\sim10^7$~K), except for the intensity peak ($>10^{9.4}$~K).  Most of the total flux was detected in the VLBA image~(Tables~\ref{table1} and \ref{table3}).  The elongated structure with high brightness temperatures in this steep-spectrum source (Table~\ref{table3}) is evidence that nonthermal jets are responsible for the observed radio emission and for making this NLS1 radio loud.  A weak counter feature was apparently seen with respect to the core, but this feature may be a mimic (Section~\ref{section:asymmetricstructure}).  We were not confident of two-sidedness in this NLS1.  The steep-spectrum core and the reasonably diffuse jet are unlikely to be caused by the strong Doppler beaming effect in this NLS1, i.e., the jets are sub-relativistic or significantly inclined.

\end{document}